\documentclass[conference, svgnames]{IEEEtran}

\usepackage{amssymb}
\usepackage{amsmath}
\usepackage{amsfonts}

\usepackage[ruled,vlined,linesnumbered]{algorithm2e}

\SetCommentSty{mycommfont}

\usepackage{booktabs} 
\usepackage{array}
\usepackage{graphicx}
\usepackage{todonotes}
\usepackage{romannum}

\usepackage{nicefrac}
\usepackage{bm}

\usepackage{booktabs} 

\usepackage{caption}

\usepackage{multirow}
\usepackage{makecell}
\usepackage[para,online,flushleft]{threeparttable}
\usepackage{adjustbox}

\usepackage{flushend}

\newcommand{\ignore}[1]{}

\usepackage{listings}
\usepackage{pifont}
\newcommand{\cmark}{\ding{51}}
\newcommand{\xmark}{\ding{55}}

\newcommand{\eg}{\textit{e.g.}\xspace}
\newcommand{\ie}{\textit{i.e.}\xspace}
\newcommand{\vs}{\textit{vs.}\xspace}

\newcommand{\squishitemize}{
 \begin{list}{$\bullet$}
  { \setlength{\itemsep}{-0pt}
     \setlength{\parsep}{2pt}
     \setlength{\topsep}{0pt}
     \setlength{\partopsep}{0pt}
     \setlength{\leftmargin}{1.0em}
     \setlength{\labelwidth}{1.0em}
     \setlength{\labelsep}{0.5em} } }

\newcounter{Lcount}
\newcommand{\squishlist}{
    \begin{list}{\arabic{Lcount}. }
   { \usecounter{Lcount}
        \setlength{\itemsep}{-0pt}
        \setlength{\parsep}{3pt}
        \setlength{\topsep}{0pt}
        \setlength{\partopsep}{0pt}
        \setlength{\leftmargin}{2em}
        \setlength{\labelwidth}{1.5em}
        \setlength{\labelsep}{0.5em} } }

\newcommand{\squishend}{\end{list}}

\usepackage{gensymb} 

\usepackage{cite}

\definecolor{Mycolor2}{HTML}{007FFF}
\newcommand\hl[1]{%
  \bgroup
  \hskip0pt\color{Mycolor2}%
  #1%
  \egroup
}

\definecolor{linkcolors}{HTML}{CC0000}

\usepackage[pagebackref=false,breaklinks=true,letterpaper=true,bookmarks=true, colorlinks, linkcolor=linkcolors, citecolor=linkcolors, urlcolor=black]{hyperref}

\usepackage[capitalise,noabbrev]{cleveref}
\crefformat{section}{\S#2#1#3} 
\crefformat{subsection}{\S#2#1#3}
\crefformat{subsubsection}{\S#2#1#3}

\begin{document}

\title{
\huge Secure Location-Aware Authentication and Communication for Intelligent Transportation Systems
}

\author{
  Nima Shoghi Ghalehshahii$^\dagger$, Ramyad Hadidi$^\dagger$, Lee Jaewon, Jun Chen, Arthur Siqueria \\
  Rahul Rajan, Shaan Dhawan, Pooya Shoghi Ghalehshahi, Hyesoon Kim\\
  \normalsize Georgia Institute of Technology\\
  \scriptsize{$\dagger$ Same Contribution} \vspace{-10pt}
  }

\maketitle

\begin{abstract}
Intelligent transportation systems (ITS) are expected to effectively create a stand-alone network for secure communication among autonomous agents. In such a dynamic and fast-changing network with high-speed agents, verifying the authenticity and integrity of messages while taking preventive action (\eg, applying brakes) within tens of milliseconds is one of the main challenges. In such a brief moment after receiving a message, the agent not only must verify the integrity and authenticity of the received message but also needs to perform extra computations to localize the sender of the message for taking appropriate action (\eg, an immediate stop warning from a vehicle in front \vs rear). In this paper, we present an inherently location-aware and lightweight authentication protocol by exploiting in situ visual localization (\ie, SLAM). In this protocol, each agent displays its public key using visual authentication beacons (\eg, QR codes). Thus, receiving agents not only can verify and authenticate the messages but also can easily localize the sender by keeping a shortlist of observed visual beacons within their visual localization system with no additional computation cost. Compared to prior work, our location-aware protocol is scalable, does not depend on any infrastructure, removes the high cost of post-message-delivery localization, and provides trustworthiness guarantees for information that are beyond the reach of each agent sensors.
\end{abstract}

\begin{IEEEkeywords}
    Intelligent Transportation Systems,
    Autonomous Agents,
    Networked Robots,
    Secure Communication
    \vspace{-10pt}
\end{IEEEkeywords}

\section{Introduction \& Motivation}
\label{sec:intro}
With recent advancements in autonomous agents, such as autonomous vehicles (AVs)\footnotemark, intelligent transportation systems (ITS) are progressing rapidly~\cite{dimitrakopoulos2010intelligent}. To enhance the safety and efficiency of road traffic, AVs can dynamically create a vehicular ad hoc network (VANET)~\cite{hartenstein2008tutorial, toor2008vehicle}. In fact, dedicated short-range communication (DSRC) channels is allocated across the world (75\,MHz in 5.9\,GHz in US and 30\,MHz in 5.9\,GHz band in Europe and Japan~\cite{fcc-dsrc, europe-dsrc}) for VANETs. Nodes in a VANET are free to move randomly and organize themselves arbitrarily with a much higher speed than conventional mobile ad hoc networks. In such a dynamic and fast-changing network, implementing a lightweight and robust security measure is crucial to ensure valid participants and integrity of messages~\cite{engoulou2014vanet, ali2019authentication}. This is because several messages in these networks are security-, safety-, and time-critical.  So, an ideal authentication method that establishes the validity of participants while delivering messages in a timely manner within a volatile environment is necessary.

\footnotetext{We interchangeably use autonomous agents and vehicles. Although our discussions are focused on vehicles, the facts are applicable to autonomous system that utilizes visual localization algorithm (\ie, SLAM).}

VANETs allow several instrumental messages that are broadly categorized under~\cite{wang2009vehicular}: (i) comfort services, such as weather information and gas stations, and (ii) safety services, such as emergency warnings, lane changing, intersection coordination, traffic-sign assistance, pre-crash warning, and forward-collision warning. We are interested in safety-related services since the messages must be processed in real-time constraints, verified to be from a trustworthy agent, and localized within the surroundings. As an example, consider a forward-collision warning that mitigates rear-end collisions by broadcasting a message to proximity vehicles of an immediate stop (due to a crash or changing traffic condition). In such a scenario, first, it is crucial to authenticate the message to ensure that it is not tampered with (\eg a bad agent tries to avoid congestion); second, it is necessary to localize the vehicle that the message originated from (\eg you should not slow down if that vehicle is behind you). Finally, the two of the aforementioned tasks must be executed within tens of milliseconds to allow the receiving vehicle to perform a preventive action (\eg applying brakes or change lanes within 100\,ms from receiving the message~\cite{lin2018architectural}).

One of the substantial potentials of VANETs is enabling shared information beyond the reach of each vehicle's sensors. Vehicles initially need to localize within their environment -- through simultaneous localization and mapping (SLAM) -- and afterwards need to update their local models as the sense changes. Since the sensing of each vehicle is constrained by its own sensors (\eg, every vehicle reaching an intersection must do all the heavy computations to detect the light), vehicles waste significant amounts of time and energy performing the same sensing and calculations to create similar information. Similar to safety services, such shared information must be also be authenticated within tens of milliseconds while ensuring that it originated from a vehicle that is physically present in the environment.

\renewcommand{\arraystretch}{1.0}
\begin{table*}[t]

\adjustbox{max width=\textwidth}{

\begin{threeparttable}
\vspace{5pt}
\centering
\caption{Authentication Methods in Vehicular Communication.}
\vspace{-0pt}
\begin{tabular}{c c || p{3.0cm} | p{3cm} | p{3cm} |c|c|c|c|c}
    \toprule
    \multicolumn{2}{c||}{\multirow{2}{*}{\textbf{Category}}}
    & \multirowcell{2}{\textbf{Short}\\\textbf{Description}}
    & \multirowcell{2}{\textbf{Pros}}
    & \multirowcell{2}{\textbf{Cons}}
    & \multirowcell{2}{\textbf{Location-}\\\textbf{Aware$^\blacklozenge$}}
    & \multirowcell{2}{\textbf{No RSU$^*$}\\\textbf{In Loop}}
    & \multirowcell{2}{\textbf{Scalibility}}
    & \multirowcell{2}{\textbf{Privacy}\\\textbf{Preserving}}
    & \multirowcell{2}{\textbf{Secure}}
    \\

    & 
    & 
    & 
    & 
    & 
    &
    & 
    & 
    &
    \\

    \midrule

    {\multirow{4}{*}{\rotatebox[origin=c]{90}{Symmetric}}}
    & \multicolumn{1}{|l||}{\multirowcell{2}{MAC Based$^1$}}
    & \multirowcell{2}{RSUs provide authentica-\\tion code each time}
    & - Fast local operation
    & - Infrastructure cost
    & \multirowcell{2}{\xmark}
    & \multirowcell{2}{\xmark}
    & \multirowcell{2}{Medium}
    & \multirowcell{2}{\cmark$^\ddagger$}
    & \multirowcell{2}{\xmark}
    \\

    & \multicolumn{1}{|l||}{}
    & 
    & - Less space overhead
    & - Susceptible to attacks
    &
    & 
    &
    &
    &
    \\ 
    \cline{3-10}
    
    & \multicolumn{1}{|l||}{\multirowcell{2}{Session-Key\\Based$^2$}}
    & \multirowcell{2}{Session key provided by\\RSU used for encryption}  
    & - Fast local operation
    & - Requires shared channel
    & \multirowcell{2}{\xmark}
    & \multirowcell{2}{\xmark}
    & \multirowcell{2}{Medium}
    & \multirowcell{2}{\cmark$^\ddagger$}
    & \multirowcell{2}{\xmark}
    \\ 

    & \multicolumn{1}{|l||}{}
    & 
    & - Lesser space overhead
    & - Susceptible to attacks
    &
    &
    &
    &
    &
    \\

    \midrule
    {\multirow{6}{*}{\rotatebox[origin=c]{90}{Asymmetric}}}
    & \multicolumn{1}{|l||}{\multirowcell{2}{Public Key\\Identification (PKI)$^3$}}
    & Authority provide keys
    & - Secure
    & - High infrastructure cost
    & \multirowcell{2}{\xmark}
    & \multirowcell{2}{\xmark}
    & \multirowcell{2}{Low}
    & \multirowcell{2}{\xmark}
    & \multirowcell{2}{\cmark}
    \\

    & \multicolumn{1}{|l||}{}
    & \& ensures security
    & - Detects bad agents
    & - Bandwidth inefficient 
    & 
    &
    &
    &
    &
    \\ 
    \cline{3-10}

    & \multicolumn{1}{|l||}{\multirowcell{2}{ID Based$^4$}}
    & Certificate management is
    & - Less bandwidth hungry
    & - Easily compromised
    & \multirowcell{2}{\xmark}
    & \multirowcell{2}{\xmark}
    & \multirowcell{2}{Medium}
    & \multirowcell{2}{\cmark$^\dagger$}
    & \multirowcell{2}{\cmark$^\dagger$}
    \\

    & \multicolumn{1}{|l||}{}
    & simplified with digital IDs
    & - Privacy preserving 
    & - Infrastructure cost
    &
    &
    &
    & 
    &
    \\

    \cline{2-10}
    & \multicolumn{1}{|l||}{\multirowcell{2}{This Work}}
    & Similar to PKI, but keys 
    & - No infrastructure needed
    & - Secure, but establishing 
    & \multirowcell{2}{\cmark}
    & \multirowcell{2}{\cmark}
    & \multirowcell{2}{High}
    & \multirowcell{2}{\cmark}
    & \multirowcell{2}{\cmark}
    \\

    & \multicolumn{1}{|l||}{}
    & are embedded as QR code
    & - Location aware 
    & trust is required
    &
    &
    &
    &
    &
    \\

    \bottomrule
    \\

\end{tabular}
\vspace{-10pt}
\begin{tablenotes}
    \item[1] \cite{asl2017synorm, rhim2012study, wang20152flip}
    \item[2] \cite{chuang2013team, wazid2017design}
    \item[3] \cite{wang2008novel, sun2010efficient, islam2018robust}
    \item[4] \cite{kamat2006identity, li2014acpn, sun2010identity, liu2014message}
    \item[*] Road side unit (RSU).
    \item[$\blacklozenge$] Integrated in the protocol itself and not requiring any extra computation afterwards.
    \item[$\dagger$] Issuer for the keys and IDs is a traffic authority (not certificate authority), which leads to both security and privacy risks. 
    \item[$\ddagger$] As long as RSUs are secure, these methods are privacy preserving. However, such an assumption is not practical.
\end{tablenotes}
\label{table:related}
\vspace{-20pt}
\end{threeparttable}

} 

\end{table*}
\renewcommand{\arraystretch}{1}
%
%
%

In this paper, we introduce a secure, lightweight, and innately location-aware protocol for vehicular communication that utilizes in situ visual localization system (\ie, SLAM) to create secure information flow between vehicles in proximity (\cref{sec:method:overview}). While SLAM has been heavily utilized in autonomous systems for localization purposes~\cite{mcmanus2013distraction, napier2012generation, wolcott2014visual, wolcott2015fast}, it has not been explored for authentication purposes. Our protocol integrates asymmetric cryptography (\cref{sec:back:pk}) and SLAM location-based visual feedback (\cref{sec:method:qr}) with inclusion of public keys as visual authentication beacons (VABs) displayed on vehicles to provide secure and lightweight authentication. Our secure protocol (\cref{sec:method:security}), unlike prior public-key-infrastructure- (PKI), ID-, MAC-, and session-key-based approaches(\cref{sec:related}), is scalable with no dependency to any infrastructure, extremely fast, location aware (\cref{sec:method:pk}), easily anonymized (\cref{sec:method:security}, \cref{sec:method:system}), and removes the high cost of post-message-delivery localization (\cref{sec:method:qr}). Our protocol has the following features:
\begin{itemize}
    \item Secure: By utilizing widespread asymmetric (public-key) cryptography.
    \item Location Aware: By integrating in situ visual localization in autonomous systems with authentication.
    \item Scalable: By eliminating dependency to infrastructure and providing trustworthiness guarantees.
    \item Fast \& Lightweight: By removing the high cost of post-message-delivery localization.
\end{itemize}

We showcase the capabilities of the protocol by simulating a VANET and observing the efficiency of the network while changing the environment (\eg, narrow street \vs highway) and the traffic level. We measure the protocol computation overhead and the effectiveness of our confirmation engine.

The rest of the paper is organized as following: \cref{sec:related} overviews  prior work, \cref{sec:back} provides necessary background, \cref{sec:method} focuses on protocol implementation, \cref{sec:method:security} verifies the security and privacy of our protocol, \cref{sec:res} presents our experimental studies, and \cref{sec:conclusion} concludes the paper.

\section{Related Work}
\label{sec:related}

A significant body of prior work has addressed construction, reliability, security, and efficiency of vehicular networks~\cite{wang2009vehicular,malhi2019security, engoulou2014vanet}. Since our work focuses on authentication, in this section, we overview current state-of-the-art authentication methods in vehicular networks, summarized in Table~\ref{table:related}.  Authentication methods are classified as either symmetric or asymmetric cryptography methods. In symmetric methods, authentication is done using the same key, whereas in asymmetric methods is done diffident, yet inter-related, keys (see~\cref{sec:back:pk} for more details). 

Symmetric cryptography methods are categorized as message-authentication-code (MAC) and session-key based. In MAC-based cryptography~\cite{asl2017synorm, rhim2012study, wang20152flip}, RSUs in an area first perform mutual authentication and key agreement with new vehicles entering the area. RSUs then use the key for future communication with the vehicles. Communication among vehicles is verified by attaching a MAC code generated using the same key. Session-key based cryptography~\cite{chuang2013team, wazid2017design} creates a new shared communication channel between any sender and the receiver by RSUs with a single key, removing the overhead of multiple trips to RSUs. Both of these methods are computationally fast because symmetric decryption is generally faster than asymmetric decryption. However, both methods rely heavily on infrastructure support. Although in session-key based approach space/communication overhead is lower than MAC-based, symmetric methods are considered not secure~\cite{malhi2019security}. From a privacy standpoint, as long as RSUs are secure, since keys are generated per session/node these methods are considered privacy preserving~\cite{malhi2019security}.

Asymmetric cryptography methods are categorized as public key identification (PKI) and ID based. Both methods use modern public-private key cryptography, in which public keys can be broadcasted. The only difference is that in PKI~\cite{wang2008novel, sun2010efficient, islam2018robust} a digital certificate is created by the certificate authority (CA) which authenticates the public key of all vehicles; whereas, in ID based~\cite{kamat2006identity, li2014acpn, sun2010identity, liu2014message} some proxy method has replaced certificate authority. The main reason is because PKI requires large bandwidth to CA that ID-based methods try to alleviate.  Both methods still need RSUs, but are secure to attacks. Moreover, PKI does not preserve privacy because the certificate used for verification can be used to identify vehicles globally~\cite{choi2011secure}.

Our protocol utilizes highly secure asymmetric cryptography, but keys are embedded as VABs, detected by in situ localization (SLAM). As a result, vehicles do not need RSUs and no infrastructure is required to scale the system (Table~\ref{table:related}). Although above methods respectively solved the security and privacy threats to different extents, they have all failed to take scalability into account. Note that moving the efficiency costs of performing complex authentication procedures to RSUs or other units transfers the load to new computational units of questionable capacity and reliability and brings up infrastructure cost issues. In fact, performing authentication without utilizing modern autonomous functionalities of modern AVs that are equipped with several sensory capabilities lead to under-utilization of these resources.

Since SLAM is inherently location aware, in our protocol, localizing vehicles does not require extra computations. Since each user has the ability to reset the VAB, tracking is impossible (\cref{sec:method:security}). Although due to the lack of infrastructure is hard to immediately verify every actor, each actor or group of actors has necessary tools (\eg, verifying behavior, trustworthiness, or databases) to establish trust with new actors (\cref{sec:method:pk:multi}). In fact, all prior work ignore the possibility that, by utilizing location-based visual feedback along with inclusion of public keys in visual clues, the need for maintaining universal systems is eliminated and the task of sender verification is reduced to a simple and computationally inexpensive task of performing a hash search of public keys against a public key repository maintained by visual scans, where the visual scans are performed as part of SLAM processes in modern autonomous systems.

\section{Background}
\label{sec:back}
\subsection{VANET Communications}
\label{sec:back:vanet}

VANET consists of two forms of communication: vehicle-to-vehicle (V2V) and vehicle-to-infrastructure (V2I). V2V communication refers to the communication of vehicles with other vehicles in their proximity usinig on-board units (OBU), either direct (\ie single-hop) or indirect (\ie multi-hop). V2I communication refers to the communication of vehicles with infrastructure, or roadside units (RSUs) such as traffic lights, traffic signs, parking meters, and installed beacons. RSUs provide infrastructure and Internet access and are operated by third parties. Some RSUs act as agents delegated by authorities (regional transportation or law enforcement). VENETs must be secure for several reasons such ensuring integrity of messages, protect privacy of users, and detecting bad actors (details in~\cref{sec:method:security}). We argue that a secure V2V communication protocol, unlike prior work, must not depend on authorities, either in the form of providing infrastructure (\eg, public key infrastructure) or verification (\eg verifying that every message is untempered with).

\subsection{Asymmetric Encryption \& Digital Signatures}
\label{sec:back:pk}

\begin{figure}[b]
    \vspace{-10pt}
    \centering
    \includegraphics[width=1.0\linewidth]{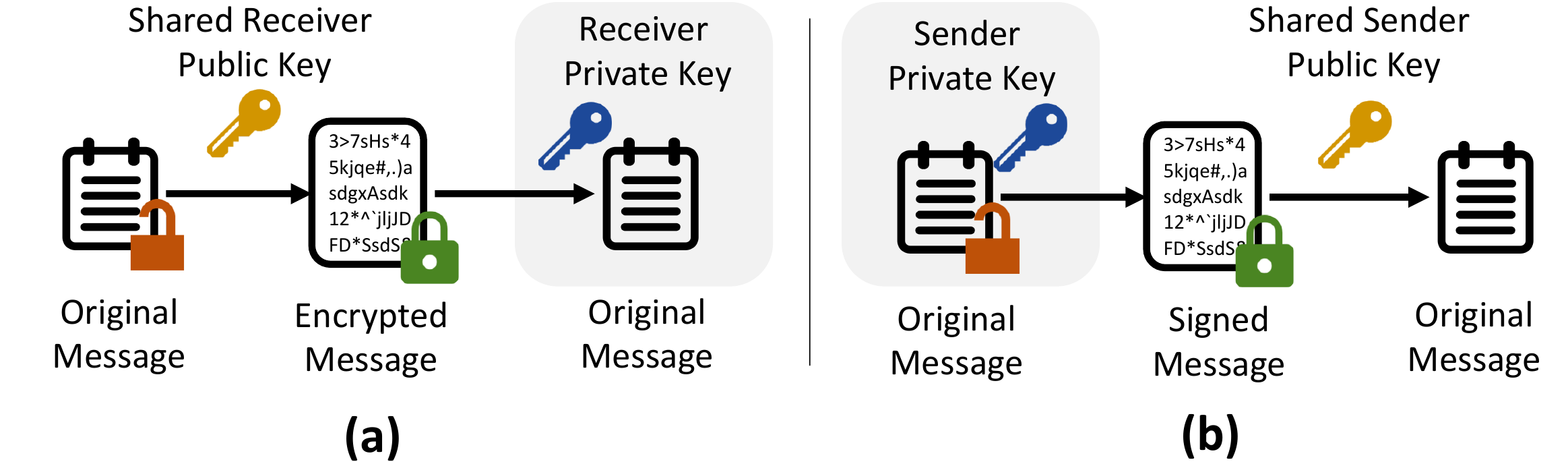}
    \vspace{-15pt}
    \caption{\textbf{Asymmetric Encryption} -- (a) Encrypting a message. (b) Signing a message to ensure authenticity.}
    \vspace{-0pt}
    \label{fig:pks}
\end{figure}

Our protocol uses widespread asymmetric cryptography (public-key cryptography), which is an essential security ingredient in modern systems, implemented in securing secure shells (SSH), virtual private networks (VPN), and encrypting disk partition~\cite{katz2014introduction}. In this cryptography, two keys are used that are tied together, private and public. Public keys can be broadcasted without any security risk, while private keys should only be known to the owner. A message intended for a receiver is encrypted with the receiver's public key that is previously broadcasted or shared. The encrypted message, however, can only be decrypted using the receiver's private key, as shown in Figure~\ref{fig:pks}a. Most systems utilize this feature of asymmetric cryptography. In our case, we use another strong feature of asymmetric cryptography, authentication. In this scenario, the sender encrypts a message with its own private key (\ie signing). A receiver can decrypt the message with the sender's public key, which proves that the message is originated from the sender since it could only have been encrypted with their private key, as shown in Figure~\ref{fig:pks}b.

\subsection{Simultaneous Localization and Mapping (SLAM)}
Simultaneous localization and mapping (SLAM)~\cite{Cadena16tro-SLAMfuture} compromises of simultaneity estimating the location of an agent while constructing a map of the environment. We focus on visual SLAM in this paper. While GPS technology identifies the general location within a map, it does not provide necessary level of precision. Hence, in addition to GPS, SLAM is a necessity for autonomous system for providing localization capability within a map with high precision~\cite{mcmanus2013distraction, napier2012generation, wolcott2014visual, wolcott2015fast}.

\section{Protocol}
\label{sec:method}
In this section, first we provide a high-level overview, then, we thoroughly discuss visual authentication beacon (VAB), on-vehicle system, and networking details. Additionally, \cref{sec:method:security} explains security and privacy features of our protocol.

\begin{figure*}[h]
    \vspace{5pt}
    \centering
    \includegraphics[width=1.0\linewidth]{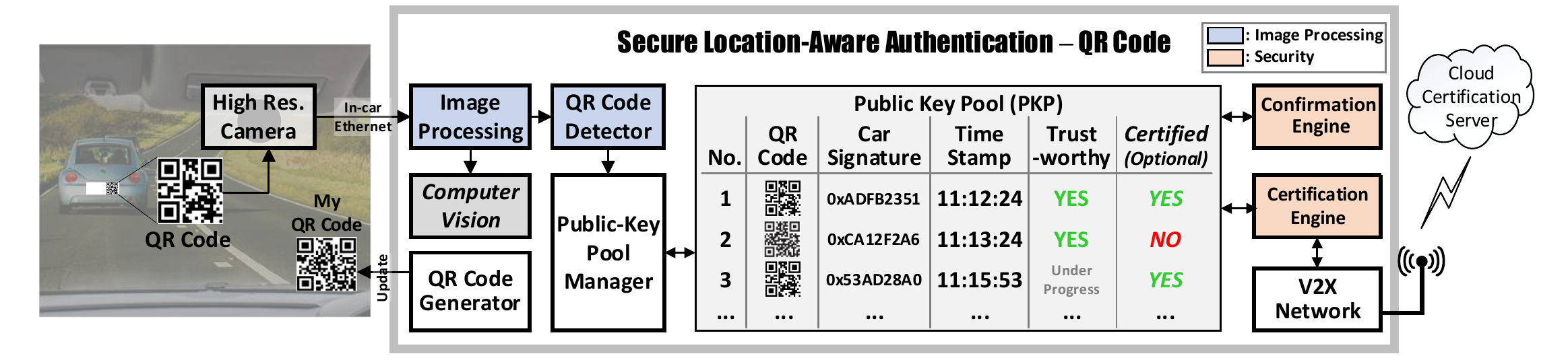}
    \vspace{-10pt}
    \caption{\textbf{On-Vehicle System Overview}}
    \vspace{-15pt}
    \label{fig:system-overview}
\end{figure*}

\subsection{Protocol Overview}
\label{sec:method:overview}

In our protocol, every vehicle has a generated public-private key pair used to sign outgoing messages. The public key is displayed on each vehicle through a VAB (\eg, QR code). Since SLAM already scans the surroundings, with a small modification, the algorithm can detect VABs and create a one-to-one mapping between objects/vehicles on its local map and their public keys. Tracking barcodes in SLAM has been extensively studied, albeit in a different context and application, by prior work~\cite{alghamdi2013indoor, zhang2015localization, rostkowska2015application}. Vehicles sign their messages with their own private key before broadcasting them in the VANET, and recipients verify the location by verifying the signature of the packet using the public key displayed on the sender's VAB. Finally, if a vehicle cannot directly verify a packet -- due to lack of visibility of the packet sender's VAB -- then it can still indirectly verify the packet if enough of its authenticated neighbors (\ie, vehicles whose VABs are visible to the vehicle) using our confirmation packet mechanism.

Our protocol provides authentication and creates a secure communication channel. Unlike prior studies, we do not rely on any infrastructure so deploying the system would not incur new costs of RSUs. The autonomous vehicles within close proximity create a secure network for communication while ensuring security and privacy. Our protocol is designed to be an applied over an abstraction of the existing VANETs. The underlying VANET and OBUs have their basic capabilities such as sending messages in a timely manner and forming networks with proximate vehicles.


\subsection{On-Vehicle System Overview}
\label{sec:method:system}

\cref{fig:system-overview} illustrates our system overview and its components. We use high-resolution cameras for SLAM and VAB sensing. When a VAB is detected, we extract that vehicle's public key and insert it, along with the vehicle's position, in our public key pool (PKP), keeping it cached for 1 minute (configurable). This table have necessary fields to track, verify, remove, or report any public key. For tracking, a unique signature that could be derived from variety of characteristics (such as a hash of unique appearances~\cite{psyllos2008vehicle} and estimated location) is created at the time of recording. Moreover, the timestamp of observation is also added. The signature and timestamp aid us in tracking the vehicle and, if necessary, removing it when no longer needed. There are also two components related to certification (optional) and confirmation engine (\cref{sec:method:pk:multi}). These components help us in detecting bad actors, reporting them to authorities (\cref{sec:method:security}), and extending the information reach of each vehicle. Additionally, for privacy purposes, VABs can be displayed on basic black-white LCD displays. Therefore, when required (\eg, end of the day), the user can generate a new QR code to protect its identity (more details in \cref{sec:method:security}).

\subsection{Digital Signatures}
\label{sec:method:qr}
\begin{figure}[t]
    \vspace{-20pt}
    \centering
    \includegraphics[width=0.8\linewidth]{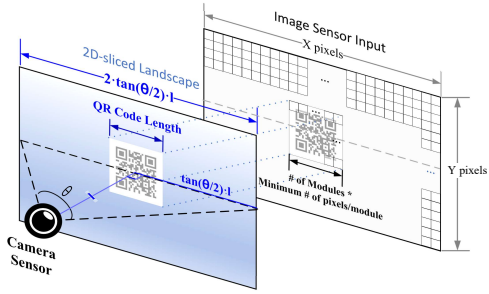}
    \vspace{-20pt}
    \caption{\textbf{QR Code Size Calculations} -- The calculation of d etecting a QR code with a on-vehicle camera.}
    \vspace{-20pt}
    \label{fig:qr_size}
\end{figure}

In our implementation, a VAB is used to visually transmit public keys to other vehicles. We experimented with barcodes such as Aztec Code, DataMatrix, and PDF-417, but we ultimately picked the QR code due to its common structure, quick recognition, and a image error correction rate as high as 30\%. A Quick Response (QR) Code is a type of 2D barcode that stores information and is recognized by digital devices. \cref{fig:qr} shows the example of a QR code. On a vehicle, a camera recognized the To recognize a QR code with a camera, mainly four factors are to be considered: distance, field of view (FOV) of camera, the resolution of the camera, and data size. In the following, we describe important factors, based on \cref{fig:qr_size}:

\begin{figure}[t]
    \vspace{-10pt}
    \centering
    \includegraphics[width=0.8\linewidth]{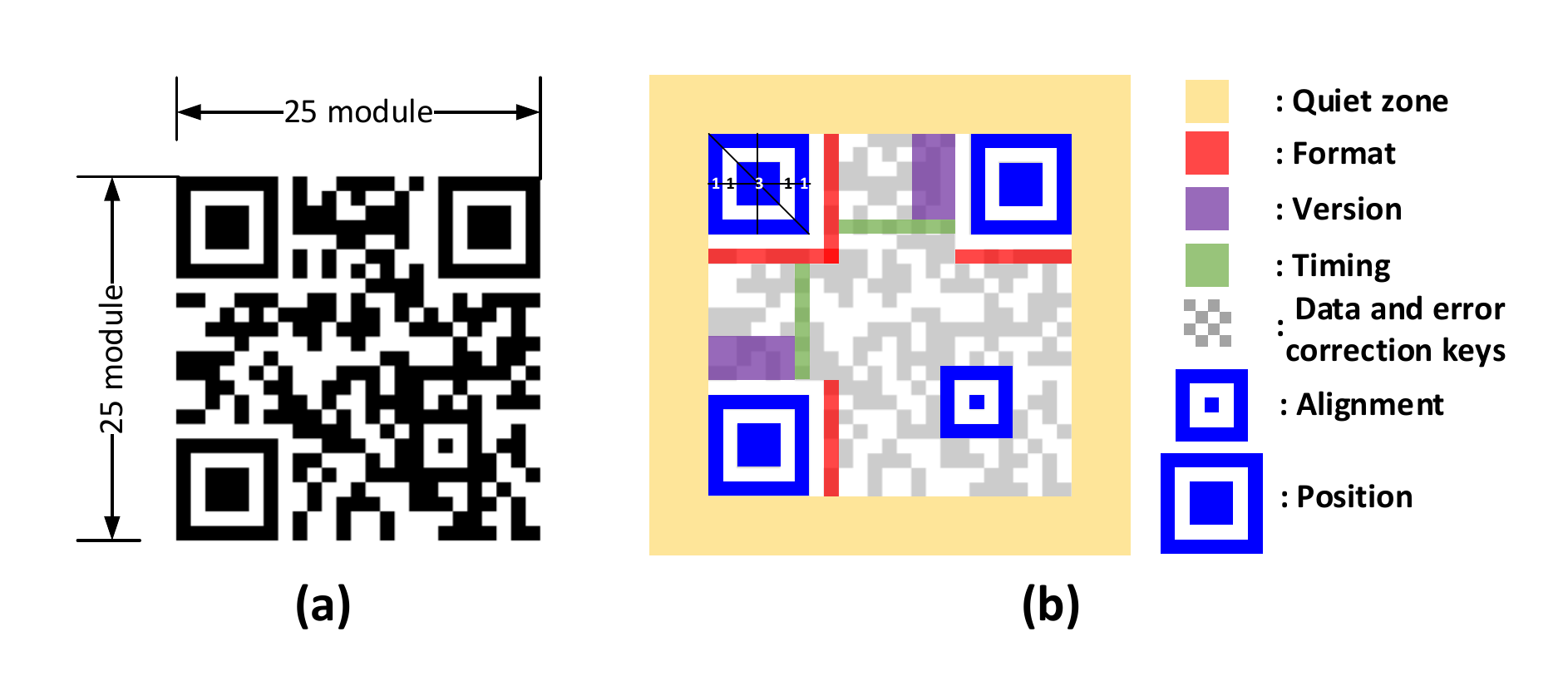}
    \vspace{-10pt}
    \caption{\textbf{QR Code Structure } -- (a) Example of QR code. (b) Components of QR code image.}
    \vspace{-15pt}
    \label{fig:qr}
\end{figure}

\noindent
$-$~\emph{Distance}, $l$: The distance between the camera and the QR code. With FOV, it determines the 2D slice of the view which directly maps to the pixels in the camera image.

\noindent
$-$~\emph{FOV}, $\theta$:  With a distance, this determines the actual length of x-axis and y-axis of the 2D slice of the view which maps to the pixels on the camera image sensor.

\noindent
$-$~\emph{Camera Resolution}, $Res_{CAM}$: This determines the actual length in real world that is mapped into a pixel. Moreover, aspect ratio of the image ($r_{CAM}$) is also a factor, commonly 3:4, but some cameras uses 2:3.

\noindent
$-$~\emph{QR Number of Modules}, $M_{QR}$: This depends on how many data points are stored in QR code, see \cref{fig:qr} as an example. In this paper, we use a public key with the size of 32 bytes. Because QR code version two consists of 25x25 pixels and need 4 quiet zone at both side, we need 33 module for a QR code image.

\noindent
$-$~\emph{QR Minimum Number of Pixels Per Module}, $P_{QR}$: Theoretically, one pixel is sufficient to capture one module in the QR code. However,  practically due to image rotation and distortion, we assume 3 pixels for recognition of a module.

Based on these metrics, we can calculate the size of the QR code, assuming that the image is located in the center of the captured image as 
\begin{equation}
\small
   \text{size}_{QR}= \frac{ 2 \cdot \tan(\frac{\theta}{2}) \cdot l}{ \sqrt{Res_{CAM}/r_{CAM}}} \cdot M_{QR} \cdot P_{QR} 
\end{equation}
When the QR code and/or the reader are moving, relative speed difference between them is also crucial. Based on our calculation with a camera resolution of 8 Megapixels (common in AVs), FOV of 30 degree, and a distance of 3 meters, the $\text{size}_{QR}=82$\,mm. By considering the fact of high-speed cars, the size would not exceed 20\,cm.

To store information in the QR code, we chose the elliptic-curve-based Ed25519 public-key signature system \cite{bernstein2012high} over other systems such as RSA due to the small key size of Ed25519 -- Ed25519's public key is 32 bytes, which is much smaller compared to the 128 bytes needed for a 1024-bit RSA public key. Such compactness is very critical because bigger key sizes lead to dense QR codes that are hard to scan. Ed25519 has some other appealing qualities, including a small signature size of 64 bytes, leading to smaller packet overhead, fast single and batch verification, and fast signing.

\subsection{Protocol Implementation}
\label{sec:method:pk}

Our protocol implementation constitutes of two main parts. First, single-hop communication that uses the broadcasting packet structure to directly send messages to nearby and visible vehicles. Second, multi-hop communication that uses confirmation packets to enable shared information, that is out of the reach of each vehicle. In the following, we describe each communication in detail besides touching on enabling a less widespread communication, point-to-point communication in VANETs. Finally, we explain and compare the overhead of our method.

\subsubsection{Single-Hop Communication (Broadcasting Packets)}
\label{sec:method:pk:single}

\begin{figure}[t]
    \vspace{5pt}
    \centering
    \includegraphics[width=1.0\linewidth]{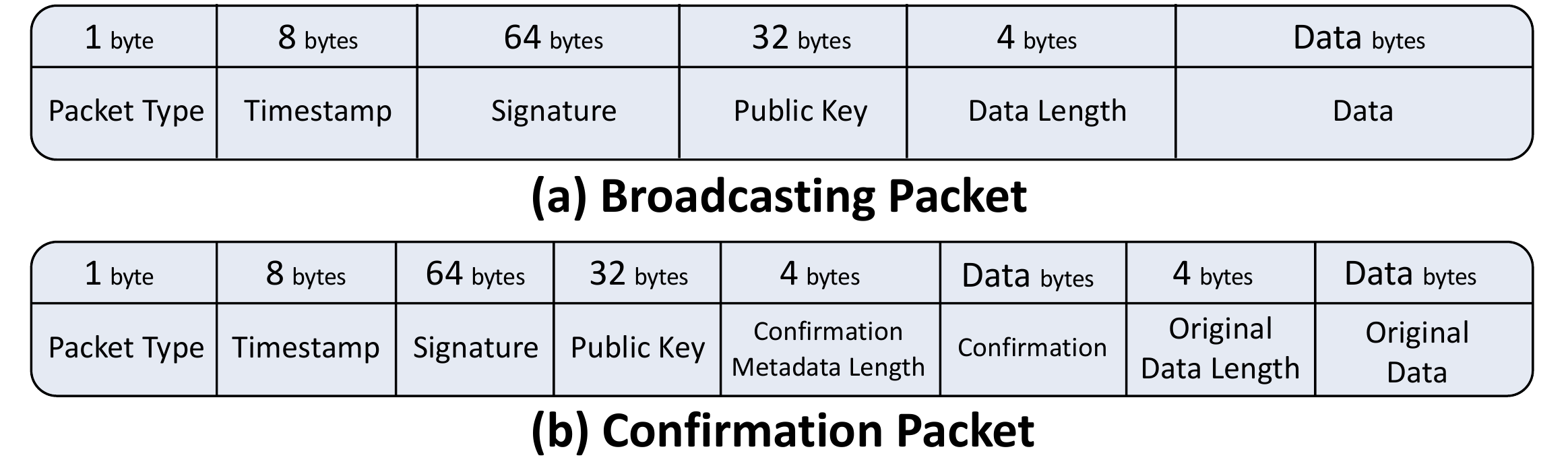}
    \vspace{-10pt}
    \caption{\textbf{Packet Structures} -- Structure of packets for broadcasting and confirmation packets.}
    \vspace{-15pt}
    \label{fig:packets}
\end{figure}

In our protocol, when a sender sends an arbitrary message, it embeds the message inside a broadcasting packet (see \cref{fig:packets}a for the packet structure) -- including a unique ID, the current timestamp, and the sender's public key in the packet header. Then, the sender generates a cryptographic signature for the packet, using its own local private key, similar to \cref{fig:pks}b, and includes this signature in the packet header. Finally, the sender broadcasts the signed packet to nearby vehicles within the VANET. When a nearby vehicle receives a packet, it runs the packet through our verification checklist, which ensures that the packet (i) has not been received yet, (ii) has not expired, (ii) is a foreign packet, (iii) is from a vehicle whose VAB has been sensed, and (iv) we can verify the packet's signature by calculating it ourselves and checking against the included signature in the packet header. Once we have verified the packet, we can also localize the sender of the packet by finding the vehicle whose VAB matches the public key of the packet's source.

\subsubsection{Multi-Hop Communication (Confirmation Packets)}
\label{sec:method:pk:multi}

To enable verified communication beyond the reach of each vehicle's sensors, receivers must be able to verify messages from other unseen senders as well. Our protocol handles this scenario through the usage of confirmation packets without RSUs. Alternatively, in the presence of RSUs, similar to prior work (See Table~\ref{table:related}), we can use certificate or traffic authorities for verification. When a vehicle verifies an incoming packet (\eg, through the single-hop scenario explained in \cref{sec:method:pk:single}), it sends out a confirmation packet, broadcasting its verification of the original packet, with the packet structure shown in \cref{fig:packets}b. The sender also includes the relative location of the original packet source in the confirmation metadata section. Vehicles outside the visibility reach of the original packet use such multiple confirmations of that packet to indirectly verify the packet, and they use the relative location metadata to localize the original sender. Figure~\ref{fig:pks} shows an example of this interaction.

We use a confirmation graph to prevent cyclic confirmations (\eg, A sends a packet; B confirms the packet; A confirms B's confirmation packet) using the confirmation engine (\cref{fig:system-overview}). The confirmation graph is a directed acyclic graph (DAG) for an original event packet whose nodes indicate vehicles and whose edges indicate a confirmation. Let $m$ be any arbitrary original message that is broadcasted, and let $\Omega_{m}$ be the confirmation graph of $m$. When a new confirmation $c$, is received, let $s$ be the sender of the confirmation, and let $s^{\prime}$ be the sender of the packet being confirmed. We start by adding a directed edge $e = (s, s^{\prime})$ to $\Omega_{m}$. Then, we detect a cyclic confirmation by performing a DFS traversal of $\Omega_{m}$ and checking if the traversal tree has any back edges. If a cycle is detected, the confirmation packet is ignored and $e$ is removed from $\Omega_{m}$.

Concretely, $m$ gets accepted if $ C(m) \geq 1 $, where $C$ is the confidence score for the received packet and is defined as
\begin{equation}
    \small
    C(m) = \sum_{p_c \in \Omega_{m}}{\frac{1}{D(p_c) + 1}}
\end{equation}
and $D$ is the depth of the packet and is defined as\footnote{Confirmations can be nested arbitrarily; therefore, $D$ is defined as a recursive function. To increase efficiency; however, $D$ can also be calculated graphically during the cycle-detection stage.}:
\[
  \small
  D(p) =  \begin{cases}
  0 & p = m \text{  (\ie $p$ is the original packet)} \\
  D(p^{\prime}) + 1 & \text{$p$ confirms $p^{\prime}$}
\end{cases}
\]
Furthermore, different confidence functions can be used for $C$ to account for different levels of trust within the system. For example, another appealing choice for $C$, shown below, prioritizes shallow confirmation (\eg 1-deep and 2-deep) over deeper confirmations:
$$
C(m) = \sum_{p_c \in \Omega_{m}}{\frac{1}{2^{D(p_c)}}}
$$

\subsubsection{Point-to-Point Communication}
\label{sec:types:pk:p2p}
Our overview of the protocol above does explicitly provide a mechanism for private and encrypted communication between two vehicles (\ie point-to-point communication between two vehicles such that other vehicles cannot eavesdrop), a less common use case. With the help of our authentication mechanism, as well as basic cryptography, we can simply extend our protocol to cover this use case. Imagine a scenario in which vehicle A sends messages to vehicle B with a requirement that vehicle C must not eavesdrop. Using the Diffie–Hellman key exchange~\cite{diffie1976new}, vehicles A and B are able to agree on a common symmetric encryption key, while the eavesdropper, C, is unable to construct the same key. Using this key, vehicles A and B are able to communicate with each other by encrypting their messages with this symmetric encryption key (\eg, using the advanced encryption standard algorithm \cite{daemen1999aes}). When A wants to send a message to B, A uses the shared key to encrypt its message. Then, B can use the same shared key to decrypt the message. Through this mechanism, we are able to implement a secure, location-aware, and point-to-point communication channel between A and B.

\subsubsection{Memory/Compute/Network Overhead}

Our protocol adds minimal memory overhead when compared to sending raw, unverified packets. Moreover, compared to all the methods in Table~\ref{table:related}, our protocol has smaller memory and computation footprint because of two main reasons: (i) we only cache a limited amount of seen VAB, and (ii) we perform the localization computations with in situ SLAM computations. For network overhead, in detail, for broadcasting packets, our protocol adds 109 bytes (see \cref{fig:packets}a) of additional memory overhead to the packet size. For confirmation packets, we incur 113 bytes of overhead plus the content of the confirmation metadata (see \cref{fig:packets}b). Additionally, we incur some network traffic overhead due to our confirmation system, as every vehicle sends out a confirmation packet for every packet received and verified. Compared to prior work, and state-of-the-art network techniques, the network overhead incurred by our protocol are at its minimum.

\section{Security \& Privacy}
\label{sec:method:security}
VANETs are mission-critical networks for autonomous systems. VANETs require not only secure communications but also the communication protocol must protect private data from various attacks. In the following, we iterate security and privacy objectives~\cite{engoulou2014vanet, malhi2019security, ali2019authentication, benyamina2019anel}. Then, we discuss common attacks and how our protocol handles them.

\noindent
$-$~\emph{Authentication:} Ensures only trusted and present users are in the network. As discussed in this section, our protocol authenticates messages with physically-present codes in the surroundings using asymmetric cryptography.

\noindent
$-$~\emph{Availability:} Ensures users can get messages at any time. Our protocol relies on current VANET technology, similar to prior work, to ensure availability.

\noindent
$-$~\emph{Integrity:} Guarantees detection of any message alteration. In our protocol, since a sender signs a message with its private key, the receiver by decrypting the message with the sender public key, can detect any alteration (\cref{sec:back:pk}).

\noindent
$-$~\emph{Privacy:} Guarantees privacy of users (\eg, the history of driving in the past day) against observers. In our protocol, each vehicle is capable of regenerating a new public-private key pair at any time to create a new identity.\footnotemark

\noindent
$-$~\emph{Real Timeliness:} Ensures messages are not computationally complex to respect real timeliness in VANETs. Our protocol uses common asymmetric encryption that has been implemented at high speeds in the chip level. Moreover, our protocol removes extra and costly localization requirements.

\noindent
$-$~\emph{Detecting Bad Actors:} Bad actors that attempt to perform attacks (\eg, a DoS attack to congest the network) should be detected and blacklisted from the network. In the following, we describe how each attack (and subsequently bad actors) are detected. After detection, the signature of the bad actor along with its key can be reported to authorities (\eg, police).

\footnotetext{Note that authorities must be able to overcome this ability, when necessary. However, integrating a backdoor in any private authentication system is debatable. Thus, similar to the current situation in traffic monitoring, we assume that authorities utilize specific technologies only accessible to them.}

\subsection*{Attacks and Countermeasures}

\noindent
$-$~\emph{Network-Level Impersonations:}
In this attack, a bad actor sends messages on behalf of another actor. This attack, executed at the network level, is not possible in our protocol. This is because our protocol enforces all messages to be signed. Thus, to impersonate, an actor must either hack the internal system of a vehicle or break the encryption, both of which are nearly impossible~\cite{katz2014introduction}.

\noindent
$-$~\emph{Replay Attacks:}
Replay attacks occur when bad actors transmit valid data that has already been signed by good actors. Our protocol prevents this attack by requiring all packets to include the current timestamp in the header of each packet and is included in the payload used to calculate the signature (\cref{fig:packets}). When a vehicle receives an incoming packet, we prevent replay attacks by: (i) If the packet is too old or too new (\ie the difference between the current time and the packet timestamp is large\footnotemark), then the packet is thrown away. (ii) If a packet is broadcasted within the allowed time-window, then there is no additional effect -- the attacker is effectively helping the target vehicle spread the original packet. (iii) If the packet is replayed but the timestamp is changed, then the signature check for the packet fails, and it will be thrown away.

\footnotetext{In our current implementation, vehicles are configured to ignore packets that are older than 5 seconds, which is configurable.}

\noindent
$-$~\emph{Eavesdropping:}
In this attack, a bad actor eavesdrops on the network traffic of a target to steal information. This protocol provides a verified public event broadcasting mechanism for vehicles. By design, all nearby vehicles should be notified of broadcasted events, and thus there is no eavesdropper protection. However, as described in \cref{sec:types:pk:p2p}, setting up a point-to-point communication channel that is safe from eavesdropping attacks is quite trivial.

\noindent
$-$~\emph{Denial of Service (DoS):}
A DoS attack occurs when an attacker floods the target with high levels of messages to make the target unresponsive. In our protocol, similar to common VANET strategies, DoS prevention and mitigation is implemented by rate-limiting combined with blacklisting/reporting to authorities.

\noindent
$-$~\emph{VAB-Level Impersonations:}
In this impersonation attack, a bad actor copies the public key of another vehicle through its VAB, pretending to be the source of those packets. While we are not able to prevent this issue, we can hold the attacker accountable. If a government-owned certificate authority is used, then police officers can verify vehicles' VABs and detect impersonators (\eg, by looking up license plates).

\noindent
$-$~\emph{Confirmation Packet Abuse:}
In this attack, a bad actor (or a group of bad actors in collusion) sends fake confirmations of illegitimate packets. In our protocol, we prevent these attacks by using the directed acyclic graph method explained in \cref{sec:method:pk:multi}. Thus, confirmation packets that would lead to confirmation cycles are ignored.

\section{Experimental Analysis}
\label{sec:res}
\subsection*{Experimental Setup}
The primary goal of our experiments is to measure the efficiency of our protocol. We emulate a group of vehicles in a VANET, laid out in different physical formations, decided by their visibility graphs. Every vehicle sends a configurable number of packets (\ie load) to the network at random time intervals. Using the MQTT protocol~\cite{hunkeler2008mqtt}, we create a network where vehicles broadcast events to their nearby vehicle. By experimenting with different visibility graphs, we are effectively able to emulate low-quality sensors and experiment with different environments for vehicles. By changing the load, we can introduce different levels of traffic. With this combination, we are able to observe and test the resiliency of the network very effectively.

\begin{figure}[h]
    \vspace{-5pt}
    \centering
    \includegraphics[width=1.0\linewidth]{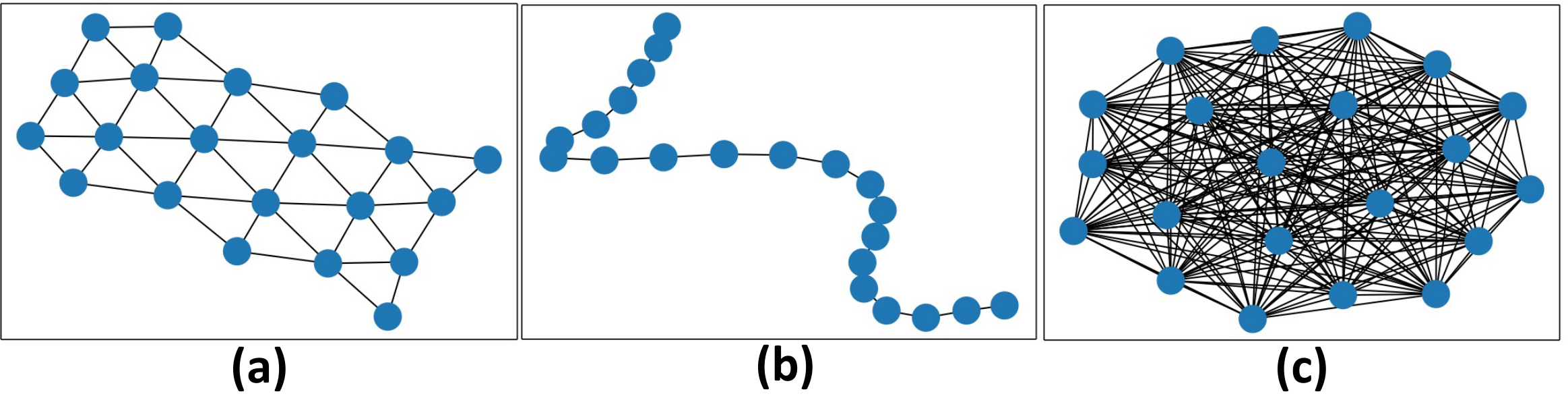}
    \vspace{-10pt}
    \caption{\textbf{Visibility Graphs} -- (a) The triangular lattice, (b) the line, and (c) the complete graphs.}
    \vspace{-5pt}
    \label{fig:graphs}
\end{figure}

\noindent
$-$~\emph{Visibility Graphs:}
Visibility graphs emulate VAB sensing in our experiments. \cref{fig:graphs} illustrates the visibility graphs used in our experiments. Every vertex in the visibility graph represents a vehicle. The existence of an edge between two vehicles means that they can see each others' VABs. Visibility graphs with higher degrees enable more routes between vehicles and thus lead to higher direct/indirect packet throughput. Each graph contains 21 nodes and is representative of the local neighborhood of a vehicle. The triangular lattice graph (\cref{fig:graphs}a) is representative of a highway with multiple lanes, where vehicles have visibility of the two vehicles in front of them. The line graph (\cref{fig:graphs}b) is representative of a narrow street where vehicles only have visibility of the vehicle that is immediately in front and behind them. The complete graph (\cref{fig:graphs}c) highlights interesting edge cases by stressing the VANET.

\noindent
$-$~\emph{Load:}
The experiments described above are ran in two different settings: low load and high load. In the low-load setting, each vehicle sends out 5 packets at random times during a time window of 1 minute. In the high-load setting, the packet count goes up to 60 during during the same time window. The high-load scenario allows us to analyze the network behavior during high-traffic situations.

\noindent
$-$~\emph{Metrics:}
We use the following metrics:
\squishitemize
    \item[$~~~~\circ$] \emph{Mean Processing Delay}, $\mu_{t}$: The average processing delay between the creation time of a packet and the reception time of a packet. This processing delay is broken down to \emph{network delay}, $t_{N}$, the delay due to network, and \emph{verification delay}, $t_{V}$, the delay due to verification.
    
    \item[$~~~~\circ$] \emph{Mean Number of Hops}, $\mu_{H}$: The average number of hops for the first verified reception of a packet (\ie, only the first successful verification is picked).
    
    \item[$~~~~\circ$] \emph{Network Reachability}, $R$: Reachability measures how well packets spread throughout the network. Concretely, $R = \frac{|P_{received}|}{|P_{sent}|}$, where $P_{sent}$ and $P_{received}$ are the sets of all outgoing original packets (not including verification packets) and verified incoming packets, respectively. \emph{Relative reachability}, $R_{\%}$, is calculated by normalizing $R$ by the highest efficiency based with that load.
\squishend

\begin{figure}[t]
    \vspace{5pt}
    \centering
    \includegraphics[width=1.0\linewidth]{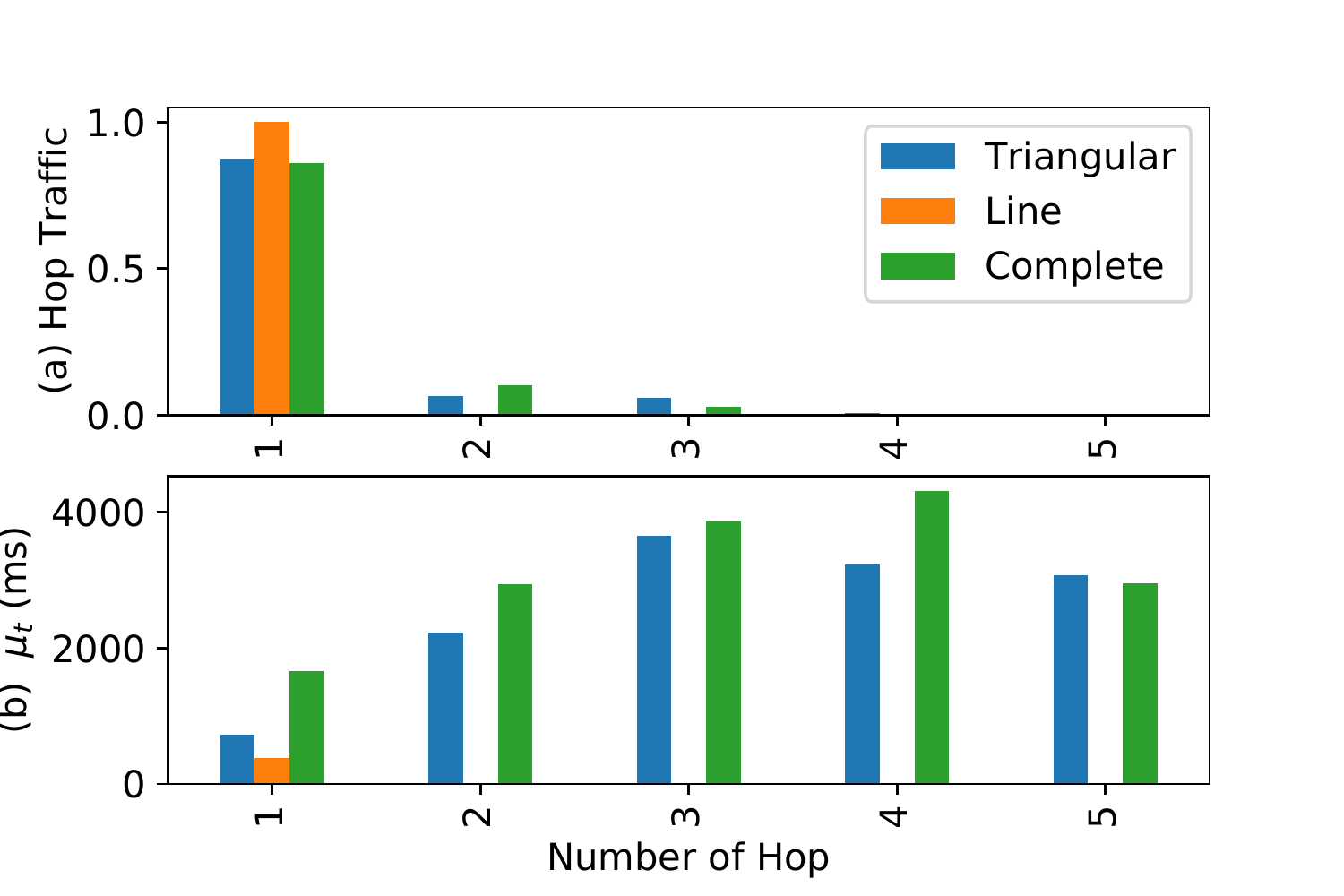}
    \vspace{-0pt}
    \caption{\textbf{Experimental Results With High Load} -- The (a) hop traffic and (b) average processing delay as functions of hop count.}
    \vspace{-10pt}
    \label{fig:graph-results}
\end{figure}

\begin{table}[b]
\vspace{-10pt}
\caption{Comprehensive Averaged Results.}
\centering
\begin{tabular}{l|l|l|l|l|l|l}
\toprule
\multicolumn{2}{c|}{\textbf{Graph}} & $\mu_t$ (ms) & $\mu_{t_{N}}$ (ms) & $\mu_H$ & $R$  & $R_{\%}$ \\ \midrule
\multirow{3}{*}{\rotatebox[origin=c]{90}{\parbox[c]{1cm}{\centering \textbf{Low Load}}}}  & \textbf{Triangular}       & 1491.71      & 1263.35      & 1.62      & 9.35      & 86\%  \\ \cline{2-7}
& \textbf{Line}             & 25.33      & 6.42        & 1.00      & 3.33       & 31\%  \\ \cline{2-7}
& \textbf{Complete}         & 1839.14      & 1596.39      & 1.26      & 10.82      & 100\% \\ \midrule
\multirow{3}{*}{\rotatebox[origin=c]{90}{\parbox[c]{1cm}{\centering \textbf{High Load}}}} & \textbf{Triangular}       & 1010.41        & 829.05    & 1.20      & 65.88       & 89\%     \\ \cline{2-7}
& \textbf{Line}             & 384.44        & 317.98     & 1.00      & 35.77       & 48\%     \\ \cline{2-7}
& \textbf{Complete}         & 1842.34        & 1598.12    & 1.17      & 74.03       & 100\%     \\ \bottomrule
\end{tabular}
\label{table:res}
\vspace{5pt}
\end{table}

\subsection{Analysis}

As shown in Table~\ref{table:res}, the processing delays of the triangular lattice graph and the complete graph are higher than that of the simple line graph in both loads. On average, we found that in single-hop scenarios, we got an \textbf{average verification delay of around 30\,ms}. For multiple-hop scenarios, this delay increased by a factor of around 650\,ms per hop. This drastic increase is not due to heavy increases in computation; instead, it is because the protocol needs multiple confirmations, and, therefore, the vehicle must wait until it receives these confirmations. Based on this, there are a few key takeaways:
\squishitemize
    \item[$\circ$] The biggest bottleneck hampering the efficiency and performance of this system is the underlying VANET. When analyzing the performance of the system in isolation (\ie, independent of the underlying network), we observe the biggest workload of the protocol to be that of indirect verifications through the confirmation system (see \cref{fig:graph-results}b).
    \item[$\circ$] This protocol has much better reachability in larger streets with more lanes as vehicles have more visibility of other vehicles' VABs in these kinds of streets. For narrow, single-lane streets, this protocol's confirmation system is practically useless.
    \item[$\circ$] The results from our complete graph experiments show that while higher visibility increases reachability, it also increases the network and verification load, leading to much higher mean processing delays.
    \item[$\circ$] \cref{fig:graph-results}a shows that confirmation levels higher than 3 are almost never utilized. Adding a hard limit on the depth of confirmation packets can heavily increase the efficiency of the protocol. This hard limit will be dependent on the confidence function chosen (See \cref{sec:method:pk:multi}).
\squishend

\section{Conclusion}
\label{sec:conclusion}
In this paper, we introduce a secure location-aware communication protocol that allows vehicles to send verified event notifications to nearby vehicles. By displaying their public key in their visual authentication beacons (VABs), such as with QR codes, vehicles can send signed messages over the VANET and have recipients verify these signed messages using asymmetric encryption. By using SLAM to detect VABs, we remove costly post-message-delivery localization. By using our confirmation engine system, we are able to send messages to vehicles beyond their VAB reach. Using this protocol, we are able to mitigate many different popular attacks, such as impersonation attacks, replay attacks, and eavesdropping attacks. Our experiments show that this protocol is effective in large systems such as streets with more lanes (\eg, highways).


\bibliographystyle{IEEEtran}
\bibliography{ms}

\end{document}